\documentstyle[11pt,AATS,epsf]{article}	

\begin{document}	

\title{Galaxy Deconstruction: Clues from Globular Clusters} 

\author{Michael J. West}
\affil{Dept. of Physics \& Astronomy, University of Hawaii, Hilo, HI 96720}
\author{}
\affil{}

\begin{abstract}
The present-day globular cluster populations of galaxies reflect 
the cumulative effects of billions of years of galaxy evolution via such 
processes as mergers, tidal stripping, accretion,
and in some cases the partial or even complete destruction of other galaxies.  
If large galaxies have grown by consuming their smaller neighbors, or by 
accreting material stripped from other galaxies, then their 
observed globular cluster systems are an amalgamation of the 
globular cluster systems of their progenitors.
Careful analysis of the globular cluster populations of galaxies can thus 
allow astronomers to reconstruct their dynamical histories.

\end{abstract}


\section{Introduction}

The origin of galaxies is one of the great outstanding problems in
modern astrophysics.  How and when did galaxies form?  How have they
evolved over time?  How does environment influence their properties?

One way to unravel the secrets of galaxy formation is by studying their 
globular cluster populations.
Most galaxies possess globular cluster systems of various richness, ranging 
from dwarf galaxies with only a handful of globulars, to supergiant 
elliptical galaxies with tens 
of thousands of globulars surrounding them 
(see Harris 1991 or van den Bergh 2000 for reviews).  
Because globular clusters are among the oldest stellar ensembles 
in the universe, 
they can provide important clues about the formation of their parent 
galaxies.

The earliest studies of globular clusters were, by necessity, limited to 
our own Galaxy and its nearest neighbors.  Over the past few decades, however, 
there has been tremendous progress in our understanding of globular cluster 
systems of other galaxies.   
One of the most important recent discoveries in 
the study of extragalactic globular 
cluster systems is that most large galaxies appear to possess two or more 
chemically distinct globular cluster populations (e.g., Gebhardt \& Kissler-Patig 1999; 
Forbes \& Forte 2000; Kundu \& Whitmore 2001). 
Some examples are shown in Figure 1, where two peaks are  
seen in the distribution of globular cluster metallicities for clusters 
associated with four large elliptical galaxies.

\begin{figure}[!t]
\plotone{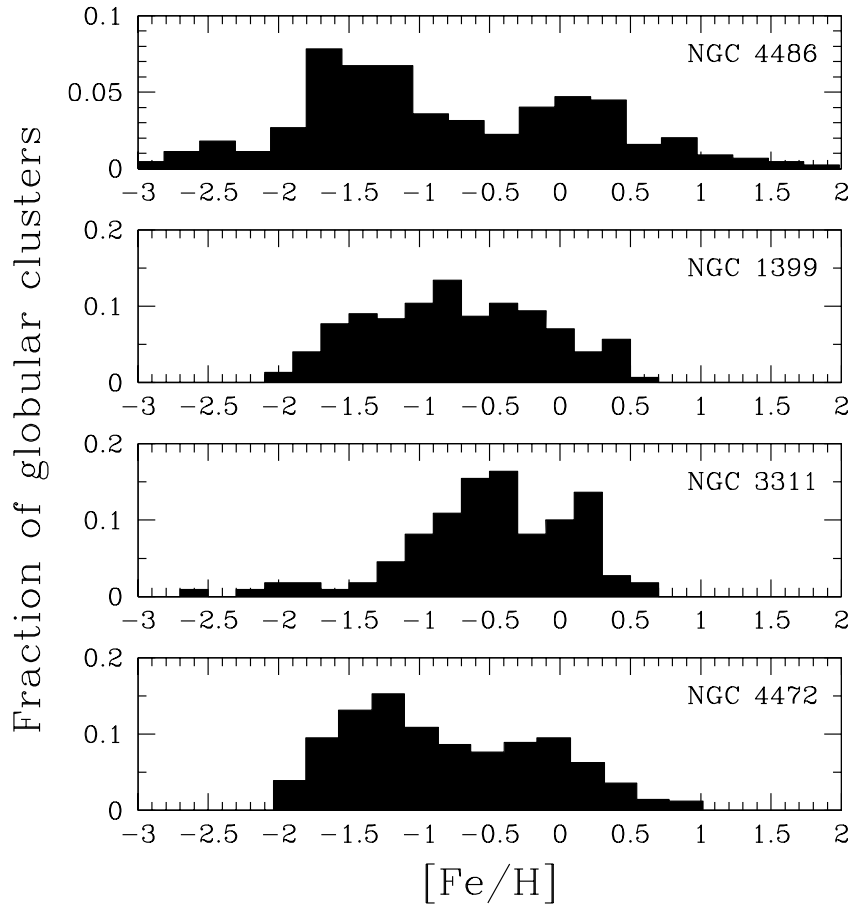}			
\caption{The observed metallicity distribution of globular cluster
systems associated with four giant elliptical galaxies.
Note the presence of two distinct peaks in most cases.  The majority of 
large elliptical galaxies studied to date exhibit bimodal globular cluster 
metallicity distributions.
\label{} }
\end{figure}

A number of different theories have been proposed to explain the origin
of these bimodal globular cluster metallicity distributions.
An obvious way to generate two or more chemically distinct globular cluster 
populations in galaxies would be through two or more bursts of globular cluster 
formation.  This might occur, for example, if 
mergers of gas-rich galaxies trigger the formation of new globulars 
(Schweizer 1987; Ashman \& Zepf 1992), resulting in 
the birth of multiple generations of globular clusters. 
Similarly, one might envision a multiphase galaxy collapse model 
in which the metal-poor globular clusters formed during the 
initial collapse of a protogalactic gas cloud, 
and the metal-rich globulars formed some time later 
(Forbes, Brodie \& Grillmair 1997; Larsen et al. 2001). 
In both of these scenarios, the metal-poor globular 
clusters surrounding galaxies such as those shown in Figure 1 would be 
their original population that formed from low-metallicity gas 
at early epochs, and the 
metal-rich globulars would have formed more recently from gas that was enriched 
by stellar evolution.

Alternatively, bimodal or multimodal globular cluster metallicity distributions 
could also arise quite naturally from galaxy mergers and/or accretion of 
globulars stripped from other galaxies {\it without needing to invoke 
the formation of multiple generations of globulars} (C\^ot\'e, Marzke, \& West 1998; 
C\^ot\'e, Marzke, West \& Minniti 2000). 
Motivation for this model came from a simple fact: 
for those elliptical 
galaxies that exhibit a bimodal 
globular cluster metallicity distribution,  
the metallicity of the 
metal-rich peak shows a clear correlation 
with parent galaxy luminosity, in the sense 
that the most luminous galaxies have the most metal-rich globulars 
(Forbes, Brodie \& Grillmair 1997; Forbes \& Forte 2001).
However, no such correlation is seen for the metal-poor peak, it   
appears to be largely independent of parent galaxy 
luminosity.  To my collaborators and I, 
this suggests that the {\it metal-rich} globular clusters are innate
to large ellipticals, and the metal-poor ones were 
added later either through mergers or accretion.

\section{Globular clusters as diagnostics of galaxy mergers}

There is no doubt that galaxy mergers have occurred frequently
throughout the history of the universe.  Figure 2 shows an 
image of a supergiant elliptical galaxy in which
the partially digested remains of several smaller galaxies 
are still clearly visible.
Many large galaxies today may have grown to their present sizes 
by devouring smaller  
companions.  If so, what becomes of the globular 
cluster populations of the galaxies that were consumed? 

\begin{figure}[!t]
\plotone{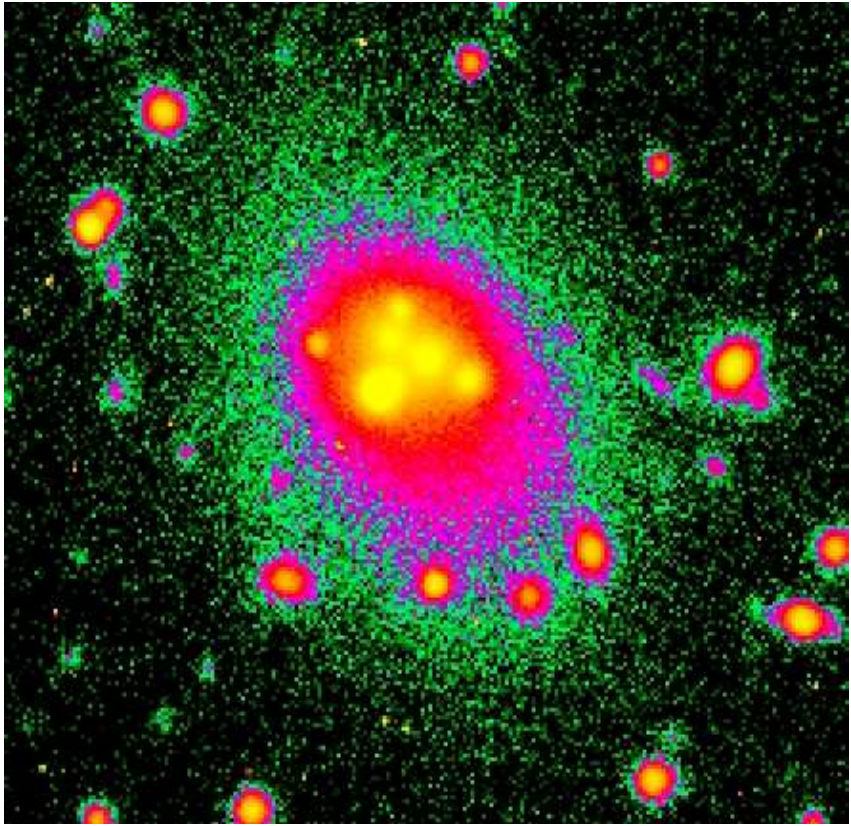}			
\caption{The brightest elliptical galaxy in the cluster Abell 3827.  Several 
smaller cannibalized galaxies are clearly evident in the central regions.
Globular clusters belonging to these galaxies are likely 
to survive the eventual disruption of their parent galaxies, and thus will 
become part of this giant elliptical.  
If most large elliptical galaxies have grown by cannibalizing smaller neighbors,
then their globular cluster populations today are composite systems that can 
provide information about the progenitor galaxies.}
\end{figure}

There is also evidence of ongoing galaxy destruction in 
rich clusters, and countless galaxies may have met their demise   
over a Hubble time (e.g., Gregg \& West 1998; Calcaneo-Rodin et al. 2000).
An example is shown in Figure 3.  
Because they are dense stellar systems, globular 
clusters are likely to survive the disruption of their parent galaxy, 
and will accumulate over time in the cores of rich galaxy clusters.  
The ongoing destruction of the moderate-sized 
elliptical galaxy shown in Figure 3, for example, will likely strew      
several hundred globulars into intergalactic space.
Some of these may
eventually be incorporated into other galaxies, a sort of
recycling on cosmic scales (Muzzio 1987). 
In particular, giant elliptical galaxies at the centers of rich 
galaxy clusters, which are observed to have enormously rich 
globular cluster populations, may have 
inherited myriad intergalactic globular clusters (West et al. 1995).

If large galaxies have grown by consuming smaller neighbors or 
by accreting material torn from other galaxies, then their 
present-day globular cluster systems are an amalgamation of 
the globular cluster systems of their victims.  
C\^ot\'e et al. (1998) and C\^ot\'e et al. (2000) 
showed that the growth of large galaxies through 
mergers or accretion will invariably be accompanied by the capture 
of metal-poor globulars, resulting in bimodal (or even 
multi-modal) metallicity distributions that are strikingly similar 
to those see in Figure 1.

\begin{figure}[!b]	
\plotone{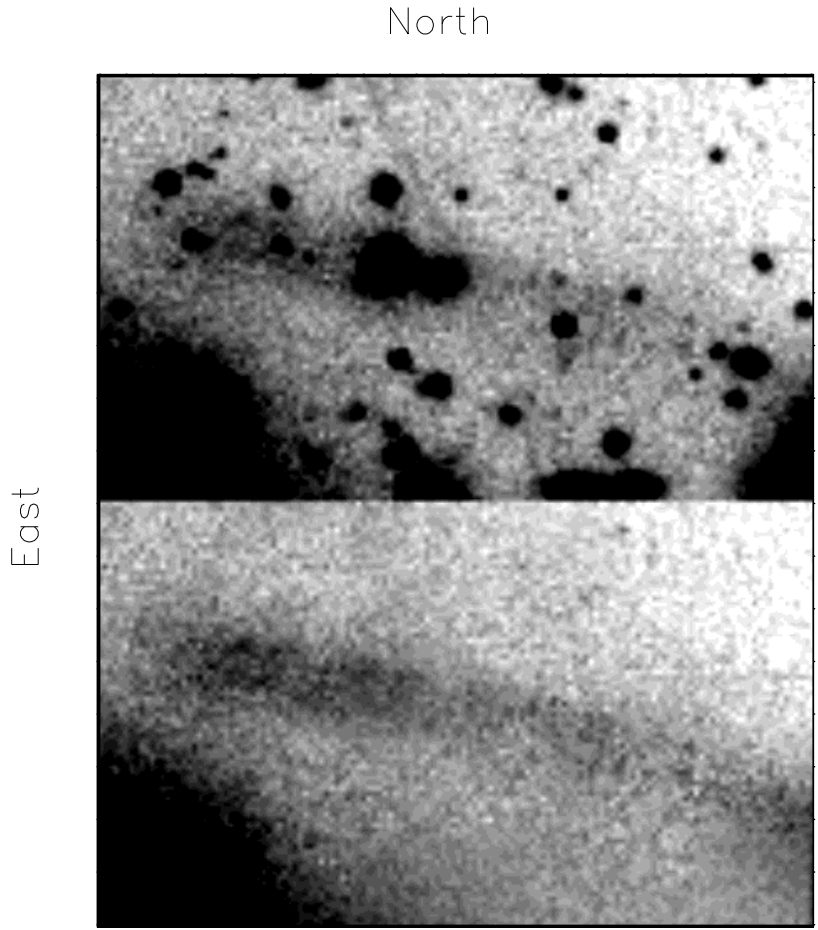}			
\caption{A tidally disrupted galaxy in the Coma cluster (from Gregg 
\& West 1998).  The top panel shows the raw image, and the bottom 
panel has been cleaned of foreground objects to highlight the 
$\sim 150$ kpc long plume 
of material.  The partial, or in some cases complete, disruption of galaxies in 
dense environments will create a 
population of intergalactic stars and globular clusters.  These freely roaming globulars 
may be accreted later by other galaxies.  
\label{} }
\end{figure}

Our prescription for building a large 
elliptical galaxy with a bimodal globular cluster metallicity distribution 
is remarkably simple:

\begin{itemize}
\item{}We assume that galaxies obey a Schechter-like luminosity function, 
as is observed.
This sets the relative numbers of galaxies of different 
luminosities that are available for merging.  

\item{}We assume that 
each galaxy is born with its own intrinsic globular cluster population,
and that the number of globulars per unit galaxy luminosity 
is constant, which is consistent with observations (Harris 1991).  
This determines how many globular clusters each galaxy 
has available to donate during mergers.

\item{}We assume, again from observations, that 
the mean metallicity of a galaxy's original globular 
cluster population increases monotonically with parent 
galaxy luminosity (C\^ot\'e et al. 1998).  
Smaller galaxies have metal-poorer globulars on average than 
larger galaxies.

\end{itemize}

Beginning with a medium-sized elliptical galaxy as a seed, 
we allow it to consume its smaller neighbors at random,
stopping after enough mergers have occurred to yield a large elliptical.
We assume that globular cluster numbers are conserved during mergers, 
so the larger galaxy gains the globulars from the smaller galaxies 
that it consumed.

\begin{figure}[!b]	
\plotone{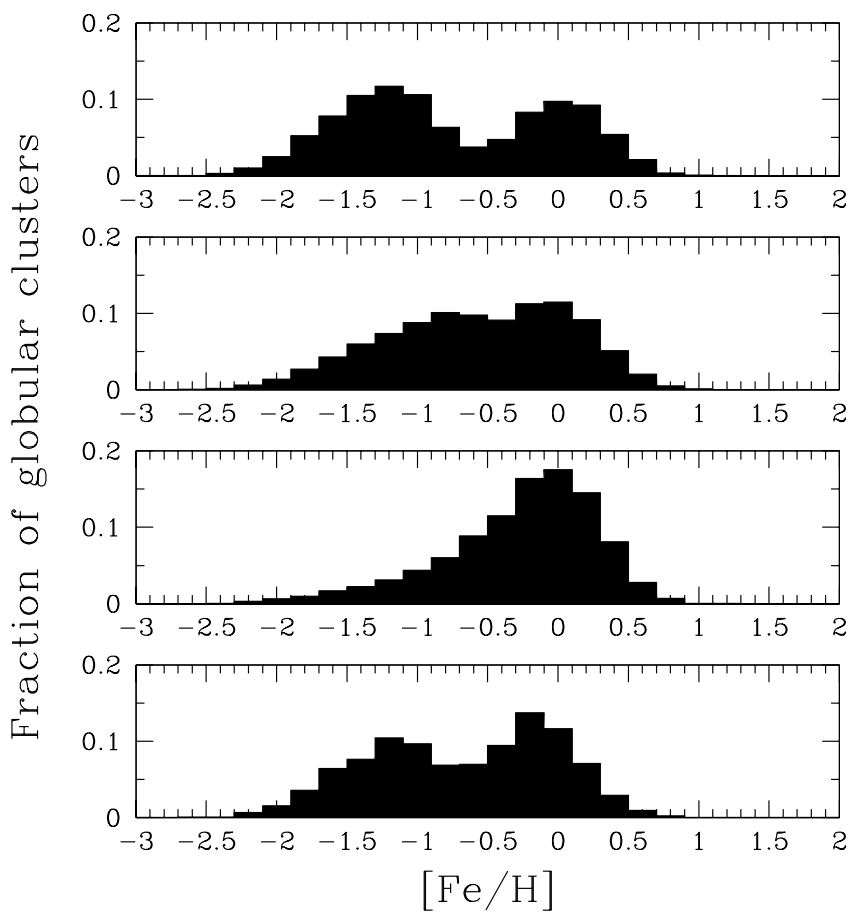}			
\caption{Results from some simulations based on 
the dissipationless merger model of C\^ot\'e, Marzke 
\& West (1998).  These simulations show that bimodal 
globular cluster metallicity distributions are easily 
produced by dissipationless galaxy mergers and accretion, without 
needing to posit the formation of multiple bursts of 
globular cluster formation.}
\end{figure}

Figure 4 shows some results of Monte Carlo simulations based on 
the C\^ot\'e et al. (1998, 2000) model.  Our simulations 
indicate that 80 to $90\%$ of large elliptical galaxies formed in this way 
exhibit bimodal (or in some cases multimodal) globular cluster distributions.  
The locations of the metal-rich and metal-poor peaks also 
agree well with observations (compare Figures 1 and 4).
In our model, the metal-rich globular clisuters of large 
ellipticals belonged to the progenitor galaxy seed, and the 
metal-poor globulars were inherited from the many smaller 
galaxies that it consumed during its growth, or by  
accretion of intergalactic globulars 
that were torn from other galaxies.  
The globulars gained from mergers or accretion 
are predominantly metal-poor because they
originate mostly in low-mass galaxies. 

It is noteworthy that unimodal globular cluster metallicity distributions  
also occur from time to time in our model.  An example can be seen 
in Figure 4.  This is not surprising, given the stochastic 
nature of the merger process.
For instance, a large elliptical could in principle be built by 
merging many small dwarf galaxies (resulting in the globular cluster 
metallicity distribution of the final merger remnant exhibiting 
a single metal-poor peak), or by merging two or three medium-sized 
galaxies (which might lead to a single metal-rich peak), 
or by merging galaxies over a 
wide range of luminosities (which yields bimodal or multimodal globular 
cluster metallicity distributions).

Unfortunately, our model seems to often be misunderstood 
or misrepresented in the literature.  For example, claims
that the C\^ot\'e et al. (1998, 2000) model is ruled out by 
the discovery of bimodality in some low-luminosity 
ellipticals (e.g., Forte et al. 2001; Kundu \& Whitmore 2001)
are completely unfounded; C\^ot\'e et al. (2000) demonstrated 
that it is quite straightforward to reproduce the bimodal 
globular cluster metallicity distribution of our own 
Galactic spheroid (which has $M_V \simeq -19.9$) as a 
consequence of accretion and mergers, and hence 
the same should be true for 
intermediate- and low-luminosity elliptical galaxies.  
Similarly, recent assertions that the location of 
the metal-poor globular cluster peak also
correlates weakly with parent galaxy luminosity is   
{\it not} ``hard to explain within the accretion/merger 
pictures'' (Larsen et al. 2001). 
A careful reading of the original C\^ot\'e, West 
\& Marzke (1998) paper would show that in fact we 
predict this very result; see Section 3.1.2 of that paper, 
which states ``there is a slight tendency for 
the brighter gE's to capture globular 
cluster populations that are more 
metal-rich than those accreted by 
their fainter counterparts (since the 
brighter gE's are able to accommodate the 
capture of more luminous intruder galaxies).''

If our model is correct, then it offers  
the exciting possibility of placing some quantitative
constraints on the number and types of mergers that galaxies 
have experienced over their lifetimes by comparing 
the relative numbers of metal-rich and metal-poor 
globulars that they possess.
C\^ot\'e et al. (1998) used this reasoning 
to conclude that M49, the most luminous elliptical 
galaxy in the Virgo cluster, must have gained roughly 
2/3 of its present luminosity by consuming other smaller Virgo 
galaxies.  More recently, we have applied these same techniques to understanding the 
formation of our own Milky Way galaxy.  C\^ot\'e et al. (2000)
showed that the present-day globular cluster system of the Milky Way 
strongly suggests that the Galaxy's spheroid   
was assembled from a large number of metal-poor protogalactic fragments.  Hence 
even a relatively low luminosity system like the Milky Way spheroid 
can and does possess a bimodal distribution of globular cluster metallicities.

\section{Where do we go from here?}

Clearly the competing theories for the origin of 
bimodal globular cluster metallicity distributions make quite different 
predictions regarding the ages of globulars.  If the metal-poor and 
metal-rich globulars surrounding large elliptical galaxies 
are the result of multiple bursts of cluster formation, then the two 
populations should have quite different ages.  If, on the other hand,
bimodal metallicity distributions can be explained by dissipationless 
merging as described above, then the metal-poor and metal-rich globulars 
should all be old.

Recently, Beasley et al. (2000) measured ages and metallicities 
of globular clusters in M49, and concluded that (within the  
sizeable uncertainties) the metal-poor and metal-rich populations 
are coeval and old.  This clearly seems to support the C\^ot\'e 
et al. (1998, 2000) picture.  However, more precise data are needed to reduce 
the uncertainties before firm conclusions can be drawn.  With that goal, 
we have obtained {\it Hubble Space Telescope} observations of 
$\sim 10^3$ globular clusters associated with the giant 
elliptical galaxy M87 in order to accurately determine their 
ages using a powerful narrow-band photometry technique.

The hypothesis that galaxies might accrete substantial numbers 
of intergalactic globular clusters also 
needs to be tested with direct observations of these objects 
to determine if they really exist and in what numbers.
Unlike many theories for the origin of globular cluster populations,
the intergalactic globulars hypothesis is easily falsifiable; if a 
significant population of intergalactic globulars is not detected 
in the cores of galaxy clusters, then this idea will have to be 
abandoned.  My collaborators and I are currently analyzing {\it HST}, Keck, 
Subaru and CFHT images that we obtained to search for  
intergalactic globulars in the Virgo, Coma and Abell 1185 galaxy 
clusters.  

There is also work to be done on the theoretical front.
The simple merger model described here is admittedly somewhat naive.  
For example, we assumed 
equal merger probabilities for all galaxies, when in reality there 
is likely to be some mass dependence of merging.  
As a step 
towards more realistic models, Frazer Pearce and I are 
collaborating in a study of the merger histories of galaxies 
that form in very high-resolution N-body cosmological simulations.  
By following 
the detailed merger histories of galaxies from high redshifts to 
the present, and inputting simple models 
of globular cluster formation at different epochs, we will be able to 
make quantitative predictions regarding 
the evolution of globular cluster metallicity distributions in
galaxies as a function of time.


\acknowledgments
I wish to thank my principal collaborators in this research, Pat C\^ot\'e, Michael Gregg, 
Ron Marzke, and Frazer Pearce.  This work was supported by NSF grant AST 00-71149. 

\newpage
\noindent
{\bf\underline{Discussion}}
\parindent=0in
\parskip=4mm

\vspace{0.1in}
{\it Michael Rich:} I support your view; in fact, in the Milky Way there are
old metal-rich globular clusters in the bulge and, as I mentioned in
my talk, when you age-date them they seem to be rather old -- at least
of order 13Gyr. We have a white dwarf distance modulus distance to 47
Tuc which gives a 13Gyr age, and then we can compare 6528 and 6553 to
that, and we think that these red globular clusters were formed along
with the Galactic bulge, very early in the history of the Galaxy. So I
think that red cluster systems have to do with
the formation of the bulge. Now, is the bulge formed in a merger
event? Maybe, the simulations show these very early mergers. On the
other hand, black hole mass is correlated with bulge velocity
dispersion and black holes are very early galaxy formation events. So
I don't know, but I would tend to support the view that the red
clusters were formed in the metal-rich spheroid.

\vspace{0.05in}
{\it West:} Thanks. 
Pat C\^ot\'e, Ron Marke, Dante Minniti 
and I published a paper just last year in which we showed that
one can account for the properties of the Milky Way spheroid and its
globular cluster system if it consumed upwards of $10^3$ little  
dwarf galaxies during its formation.

\vspace{0.05in}
{\it Hugh Harris:} The clusters are usually found more widely distributed
spatially than the halo light in their galaxy. How do you interpret
that?

\vspace{0.05in}
{\it West:} We think that accretion, not just mergers, 
has probably played an important role 
by stripping globular clusters from some galaxies and 
then adding these to  
the outer regions of others, especially giant ellipticals in the
centers of rich clusters.  This would 
explain why some of large
ellipticals, like M87 for example, have far more globulars per unit
galaxy luminosity than expected, and why they have very extended 
distributions.   

\vspace{0.05in}
{\it Alan Stockton:} The problem with making large bulges or large
ellipticals from a lot of small things, of course, is the
luminosity-color relation, or luminosity-metallicity relation. So you
indicated in most cases you favor starting with a fairly large
progenitor and then putting a bunch of small things on it.  Do you 
still run into any problem with the observed dispersion in that
relation?

\vspace{0.05in}
{\it West:}  One can imagine making a large elliptical in
many different ways, either from the merger of many small dwarfs, in
which case the elliptical should be pretty blue as well, or from a
number of bigger progenitors which would have had redder colors. 
The existence of the observed luminosity-color relation 
for galaxies suggests that in most cases large ellipticals
can't have formed just from the merger of many dwarfs -- that
probably wouldn't work, you're right.
It suggests that most large ellipticals must have started from 
fairly large progenitor seeds.  As this seed galaxy consumes 
smaller neighbors, this will dilute the luminosity-color 
relation, but won't necessarily obliterate it.

\end{document}